\documentclass[aps,prc,showpacs,twocolumn,preprintnumbers,floatfix, tightenlines]{revtex4}
\usepackage{amsmath,amssymb,amsfonts}
\usepackage{hyperref}
\usepackage{ulem}
\usepackage{graphicx}
\usepackage{color}
\usepackage{longtable}

\allowdisplaybreaks

\begin{document}

\title{Sensitivity of symmetry energy elements of nuclear matter to the 
\\properties of neutron rich systems}
\author{C. Mondal}
\address{Saha Institute of Nuclear Physics, 1/AF Bidhannagar, Kolkata
{\sl 700064}, India}
\author{B. K. Agrawal}
\email {bijay.agrawal@saha.ac.in}
\address{Saha Institute of Nuclear Physics, 1/AF Bidhannagar, Kolkata
{\sl 700064}, India}
\author{J. N. De}
\address{Saha Institute of Nuclear Physics, 1/AF Bidhannagar, Kolkata
{\sl 700064}, India}
\author{S. K. Samaddar}
\address{Saha Institute of Nuclear Physics, 1/AF Bidhannagar, Kolkata
{\sl 700064}, India}

\begin{abstract}
The sensitivity of nuclear symmetry energy elements at the saturation
density to the binding energies of ultra neutron-rich nuclei (neutron to
proton ratio $\sim$ 2) and the maximum mass of neutron star is explored
within a relativistic mean field model.  Values of the interaction
parameters governing the isovector strengths and the symmetry elements
are determined in tighter bounds.  Assessments based on the sensitivity
matrix reveal that the properties of extreme neutron-rich systems play
a predominant role in narrowing down the uncertainties in the various
symmetry energy parameters.  The calculations are extended over a wide
range of nuclear matter density and the results are discussed.

\end{abstract}
\maketitle


\section{Introduction}
The binding energies of atomic nuclei are the most accurately determined
data in nuclear physics. All informations about nuclear interactions
are firmly entrenched in them. Ideas have been proposed in recent years to
establish precision values of isovector indicators in nuclear interactions
(like the volume and surface symmetry energy coefficients $C_v^0$
and $C_s$ of nuclei, the density slope parameter $L_0$ of symmetry
energy or the neutron skin $\Delta r_{np}$ of heavy nuclei) from nuclear
masses. Using the nuclear droplet model designed to explain nuclear
masses, from correlation systematics of $\Delta r_{np}$ of $^{208}$Pb
with nuclear isospin, Centelles {\it et al} \cite{Centelles09} get
$L_0 = 75 \pm 25$ MeV. Fitting few thousand observed nuclear masses
within the finite range droplet model (FRDM) \cite{Moller12} M\"oller
{\it et al} find $C_v^0$ to be $32.5\pm0.5$ MeV, $L_0$ comes out to be
$70\pm15$ MeV. From double differences of symmetry energies estimated
from the nuclear masses, Jiang {\it et al} \cite{Jiang12} get $C_v^0
= 32.10\pm0.31$ MeV and the surface symmetry energy coefficient
$C_s$ is determined to be $58.91\pm1.08$ MeV. These informations
coupled with microscopic calculations on nucleon distributions
\cite{Agrawal12,Agrawal13,Liu13} in a heavy nucleus like $^{208}$Pb give
$L_0 = 59\pm13$ MeV and the neutron skin $\Delta r_{np}$ of $^{208}$Pb
to be $\sim 0.195\pm0.021$ fm. With progression in time, analyses based
on nuclear masses seem to contain the fluctuations in the symmetry
observables better \cite{Moller12,Jiang12,Danielewicz09,Liu10,Wang13},
better compared to those obtained from isoscaling \cite{Shetty07},
nuclear emission ratios \cite{Famiano06}, isospin diffusion \cite{Li08},
giant dipole polarizability \cite{Roca-Maza13}, quadrupole resonances
\cite{Roca-Maza13a} or from astrophysical evidences \cite{Steiner12}.

Microscopic mean-field models, both non-relativistic and relativistic with
energy density functionals (EDF) parametrized to give best fits to nuclear
masses and also to some other selective isoscalar and isovector-sensitive
observables, however, do not show such constraints in the values of
$C_v^0$ or $L_0$ \cite{ Dutra12, Dutra14, Niksic08, Zhao10,Dong12}. They
have much wider variations. This looks like a puzzle. As an example, the
isovector sensitive binding energy difference $\Delta B$ of $^{132}$Sn
and $^{100}$Sn shows nearly no correlation between $\Delta B$ and
$\Delta r_{np}$ for $^{132}$Sn for different set of Skyrme interactions
\cite{Brown00}. The binding energy difference in different EDFs are
reproduced fairly well, but $\Delta r_{np}$ shows a wide variation.

In a recent calculation \cite{Mondal15} in the relativistic mean
field model (RMF), attempts were made to understand this seeming
paradox. The binding energy difference between four pairs of nuclei,
namely ($^{68}$Ni - $^{56}$Ni), ($^{132}$Sn - $^{100}$Sn), ($^{24}$O -
$^{16}$O) and ($^{30}$Ne - $^{18}$Ne) calculated in 14 different RMF
models when plotted against the computed values of $\Delta r_{np}$ of
$^{208}$Pb in the same models shows increasingly high correlation with
increasing asymmetry. The asymmetry $\delta$ is defined as $\delta =
(N-Z)/A$ where $N$ and $Z$ are the neutron and proton numbers in a nucleus
of mass number $A$. This is suggestive of the fact that isovector part of
the nuclear interaction would be better constrained if nuclei of higher
asymmetries are included in the binding energy systematics. A covariance
analysis displaying correlation among physical observables makes this
fact more revealing.  That the choice of very neutron-rich isotopes is
more rewarding for the searching of isovector signatures was also found
recently by Chen and Piekarewicz \cite{Chen15}.

The present communication is a follow-up of our earlier
paper \cite{Mondal15} and to see if the conclusions so drawn in
Refs. \cite{Mondal15,Chen15} stand a broader systematic analysis.  The
sensitivity of symmetry energy elements of nuclear matter at saturation
density to different properties of highly neutron rich systems are
analyzed here in detail.  For this, we have chosen an expanded data set
of nuclear masses with inclusion of nuclei of very high isospin ($\delta
> 0.3$). Since neutron star is composed mainly of  nuclear matter   of
extreme isospin, we have also included the  observed maximum  mass of
neutron star as an element in the fitting protocol.  The analysis is
also extended over a wide range of nuclear matter density.

\begin{table*}[t]
\caption{\label{tab1}
Optimum values of the parameters for the models SINPB and SINPA,
uncorrelated (upper line) and correlated (lower line) errors on them
are given in the parentheses. Mass of the $\sigma$ meson ($m_\sigma$)
is given in units of MeV.  The masses of $\omega$ and $\rho$ mesons are
kept fixed to $m_\omega$= 782.5 MeV and $m_{\boldsymbol\rho}$= 763 MeV
and nucleon mass is taken to be $M$= 939 MeV.
} 
 \begin{ruledtabular}
\begin{tabular}{ccccccccc}
Name  & $g_{\sigma}$ & $g_{\omega}$ & $g_{\boldsymbol\rho}$ & ${\kappa_3}$ & ${\kappa_4}$ 
& ${\eta_{2\boldsymbol\rho}}$ & $\zeta_0$ & $m_{\sigma}$ \\
\hline
SINPB &-10.6007 & 13.8767 & 10.613 & 1.4868 & -0.802 & 13.487 & 5.467 & 
493.850\\
& (0.0001) & (0.0002) & (0.036) & (0.0002) & (0.002) & (0.312) & (0.003) & 
(0.006) \\
& (0.14) & (0.24) & (1.29) & (0.19) & (1.15) & (12.26) & (0.45) & (4.98)\\
SINPA &-10.6292 & 13.8532 & 12.831 & 1.5375 & -1.190 & 38.179 & 5.363 & 
495.394 \\
& (0.0001) & (0.0002) & (0.043) & (0.0002) & (0.002) & (0.521) & (0.003) & 
(0.006) \\
& (0.16) & (0.33) & (0.82) & (0.06) & (0.47) & (11.92) & (0.45) & (3.86)\\
\end{tabular}
\end{ruledtabular}
\end{table*}

The paper is organized as follows. The effective Lagrangian density for
the RMF model employed in the present work is briefly outlined in Sec. II.
The results obtained by optimizing the objective function for different
sets of fit-data and the analysis pertaining to the sensitivity of various
symmetry energy parameters to the properties of neutron rich systems are
discussed in Sec. III.  The summary and conclusions are given in Sec. IV.

\section{Effective Lagrangian}
The effective Lagrangian density for the RMF model used in this present
work is similar to that of FSU \cite{Todd-Rutel05, Furnstahl97, Boguta77,
Boguta83}.  Its interaction part is given by,
\begin{eqnarray} 
\label{Lagrangian} 
{\cal L}_{\it int}=&&\overline{\psi}\left [g_{\sigma} \sigma -\gamma^{\mu} \left (g_{\omega }
\omega_{\mu}+\frac{1}{2}g_{\mathbf{\boldsymbol\rho}}\boldsymbol\tau .
{\boldsymbol\rho}_{\mu}+\frac{e}{2}(1+\tau_3)A_{\mu}\right ) \right ]\psi\nonumber
\\ &&-\frac{{\kappa_3}}{6M}
g_{\sigma}m_{\sigma}^2\sigma^3-\frac{{\kappa_4}}{24M^2}g_{\sigma}^2
m_{\sigma}^2\sigma^4+ \frac{1}{24}\zeta_0
g_{\omega}^{2}(\omega_{\mu}\omega^{\mu})^{2}\nonumber
\\ &&+\frac{\eta_{2\boldsymbol\rho}}{4M^2}g_{\omega}^2m_{\boldsymbol\rho
}^{2}\omega_{\mu}\omega^{\mu}\boldsymbol\rho_{\nu}\boldsymbol\rho^{\nu}.  
\end{eqnarray}
It comprises the conventional Yukawa couplings between the
nucleonic field $\psi$ and the mesonic fields $\sigma$, $\omega$ and
$\boldsymbol\rho$ with coupling constants $g_{\sigma}$, $g_{\omega}$
and $g_{\boldsymbol\rho}$, respectively. The parameter $g_{\sigma}$
stands for the long-range attractive nature of the nuclear force. Its
repulsive nature at short ranges is taken care of by the parameter
$g_{\omega}$. The cubic and quartic self couplings of $\sigma$ meson are
characterized by the parameters $\kappa_3$ and $\kappa_4$, respectively.
The parameter $\zeta_0$ symbolizes the strength of the quartic self
coupling of $\omega$ meson. All these self-couplings render an added
softening in the equation of state (EoS) of symmetric nuclear matter. The
density dependence of symmetry energy of nuclear matter is governed by the
coupling constants $g_{\boldsymbol\rho}$ and $\eta_{2\boldsymbol\rho}$;
$\eta_{2\boldsymbol\rho}$ represents the strength of the cross-coupling
between $\omega$ and $\boldsymbol\rho$ mesons. The parameter
 $\eta_{2\boldsymbol\rho}$ provides some added flexibility in
fitting the ground state properties of some standard doubly magic
spherical nuclei and yet allowing the value of the neutron skin $\Delta
r_{np}$ of $^{208}$Pb to vary over a wide range \cite{Furnstahl02, Sil05}.
The strength of the electromagnetic interaction between the protons is
described by the coupling constant $e$.

\section{Results and Discussions}
To analyze the sensitivity of the symmetry energy elements of nuclear
matter to highly neutron rich systems, two different RMF models are
constructed, one with inclusion of few ultra neutron-rich systems
in the fitted data and another without them. A comparative study on the
nuclear matter properties of these two models is executed in detail. We
performed a comprehensive study to find the isovector signatures from
those highly neutron rich systems from different perspectives. At the end,
different properties of nuclear matter beyond saturation are critically
examined.

\subsection{The RMF models SINPB and SINPA}
The values of the parameters ({\bf p}) which describe the EDF of the RMF
model (Eq. (\ref{Lagrangian})) are obtained by optimizing the
objective function $\chi^2({\bf p})$ defined as,
\begin{equation} 
\chi^2({\bf p}) =\frac{1}{N_d - N_p}\sum_{i=1}^{N_d} \left (\frac{ \mathcal{O}_i^{exp} 
- \mathcal{O}_i^{th}({\bf p})}{\Delta\mathcal{O}_i}\right )^2.
\label {chi2} 
\end{equation} 
Here, $(N_d - N_p)$ is the degrees of the freedom of the system given
by the difference between the number of experimental data points
$N_d$ and number of fitted parameters $N_p$.  The experimental and the
corresponding theoretically obtained value of an observable are given by
$\mathcal{O}_i^{exp}$ and $\mathcal{O}_i^{th}({\bf p})$, respectively.
$\Delta\mathcal{O}_i$ is the adopted error of an observable.  The minimum
value of the objective function $\chi_0^2$ corresponds to the value
of $\chi^2$ at ${\bf p_0}$, ${\bf p_0}$ being the optimal values of
the parameters.

After optimization \cite{Bevington69}, variance on a quantity
$\mathcal{A}$, ${\overline{\Delta \mathcal{A}^2}}$ can be evaluated as,
\begin{equation}
\overline{\Delta \mathcal{A}^2}=\sum_{\alpha\beta}\left(
\frac{\partial \mathcal{A}}{\partial
\rm{p}_{\alpha}}\right)_{\bf p_0}\mathcal{C}_{\alpha\beta}^{-1}\left(\frac
{\partial \mathcal{A}}{\partial \rm{p}_{\beta}}
\right)_{\bf p_0}, 
\label{deltaab}
\end{equation}
where $\mathcal{A}$ could be parameters as
well as observables.
Here $\mathcal{C}_{\alpha\beta}^{-1}$ is an element of inverted curvature
matrix given by,
\begin{equation}
 \mathcal{C}_{\alpha\beta}=\frac{1}{2}\Big(\frac{\partial^2 \chi^2(\mathbf{p})}
{\partial {\rm {p}_{\alpha}}\partial {\rm{p}_{\beta}}}\Big)
_{\mathbf{p}_0}.
\label{Cmatrix}
\end{equation}
From Eq. (\ref{deltaab}) one can see that, the diagonal element
$\mathcal{C}_{\alpha\alpha}^{-1}$ of the inverted curvature matrix is the
variance on the parameter $\rm{p}_{\alpha}$.
\begin{table}[h]
\caption{\label{tab2}
Various observables $\mathcal{O}$, adopted errors on them
$\Delta\mathcal{O}$, corresponding experimental data (Expt.) and their
best-fit values for SINPB and SINPA. $B$ and $r_{ch}$ corresponds to
binding energy and charge radius of a nucleus, respectively and $M_{max}^{NS}$
is the maximum mass of neutron star (NS). Values of $B$ are given in
units of MeV and $r_{ch}$ in fm.  $M_{max}^{NS}$ is in units of Solar Mass
($M_{\odot}$).}
 \begin{ruledtabular}
\begin{tabular}{cccccc}
& $\mathcal{O}$ &$\Delta\mathcal{O}$ & Expt. &  
SINPB & SINPA \\
\hline
$^{16}$O & $B$ &4.0 & 127.62 &  127.78 & 128.35\\
& $r_{ch}$ &0.04 & 2.699 &  2.704 & 2.696\\
$^{24}$O & $B$ &2.0 & 168.96 & - & 169.28\\
$^{20}$Ne& $B$ &4.0 & 160.64 & - & 155.89\\
$^{30}$Ne& $B$ &3.0 & 211.29 & - & 214.37\\
$^{24}$Mg& $B$ &3.0 & 198.26 & - & 195.87\\
$^{36}$Mg& $B$ &2.0 & 260.78 & - & 261.68\\
$^{40}$Ca& $B$ &3.0 & 342.05 &  343.19 & 343.66\\
& $r_{ch}$ &0.02 & 3.478 &  3.460 & 3.452\\
$^{48}$Ca& $B$ &1.0 & 416.00 &  415.27 & 415.47\\
& $r_{ch}$ &0.04 & 3.477 &  3.437 & 3.437\\
$^{54}$Ca& $B$ &2.0 & 445.37 &  445.63 & 443.79\\
$^{58}$Ca& $B$ &2.0 & 454.43 & - & 456.33\\
$^{56}$Ni& $B$ &5.0 & 483.99 &  483.38 & 484.34\\
& $r_{ch}$ &0.18 & 3.750 &  3.700 & 3.686\\
$^{68}$Ni& $B$ &2.0 & 590.41 &  592.86 & 592.97\\ 
$^{78}$Ni& $B$ &2.0 & 641.78 &  642.10 & 641.59\\ 
$^{90}$Zr& $B$ &1.0 & 783.90 &  783.02 & 783.20\\
& $r_{ch}$ &0.02 & 4.269 &  4.266 & 4.264\\
$^{100}$Sn& $B$ &2.0 & 825.30 & 828.11 & 827.93\\
$^{116}$Sn& $B$ &2.0 & 988.68 & 987.45 & 987.32\\
& $r_{ch}$&0.18 & 4.625 &  4.620 & 4.622\\
$^{132}$Sn& $B$ &1.0 & 1102.84 & 1103.28 & 1103.40\\
& $r_{ch}$&0.02 & 4.709 &  4.706 & 4.710\\
$^{138}$Sn& $B$ &2.0 & 1119.59 &  1118.65 & 1117.05\\
$^{144}$Sm& $B$ &2.0 & 1195.73 &  1196.00 & 1195.67\\
& $r_{ch}$&0.02 & 4.952 &  4.955 & 4.955\\
$^{208}$Pb& $B$ &1.0 & 1636.43 &  1636.38 & 1636.57\\
& $r_{ch}$&0.02 & 5.501 &  5.528 & 5.530\\
\hline
NS& $M_{max}^{NS}$ &0.04 & 2.01 & - & 1.98\\
\end{tabular}
\end{ruledtabular}
\end{table}

To explore the information content of some of the highly neutron rich
systems, two models, namely, SINPB and SINPA are constructed.  In SINPB
binding energies ($B$) and charge radii ($r_{ch}$) of some standard
set of nuclei across the whole nuclear chart are taken as fit-data
(see Tab. \ref{tab2}).  The binding energies of $^{54}$Ca, $^{78}$Ni and
$^{138}$Sn nuclei having somewhat larger asymmetry ($\delta\sim$ 0.26 -
0.28) are also included in the fitting protocol. The model SINPA includes
some highly asymmetric nuclei, namely, $^{24}$O, $^{30}$Ne, $^{36}$Mg and
$^{58}$Ca ($\delta>$ 0.3) in addition to the data set used in the base
model SINPB. SINPA also contains the symmetric $^{20}$Ne and $^{24}$Mg
nuclei and the observed maximum mass of neutron star $M_{max}^{NS}$
as fit-data.

In Table \ref{tab1}, the optimal values of the parameters {\bf p$_0$}
for SINPB and SINPA are given along with the uncorrelated and correlated
errors on them. Correlated or uncorrelated errors can be calculated from
Eq. (\ref{deltaab}) by including or excluding the contribution from the
off-diagonal elements of curvature matrix $\mathcal{C}_{\alpha\beta}$,
respectively. One can observe that the correlated errors (lower
lines inside the parentheses in Table \ref{tab1}) on the parameters
$g_{\boldsymbol\rho}$, ${\kappa_3}$, ${\kappa_4}$ and $m_{\sigma}$
decreased by a noticeable amount in SINPA in comparison to SINPB. For
the parameters $g_{\sigma}$, $\zeta_0$ and ${\eta_{2\boldsymbol\rho}}$
the errors are almost the same for both the models and for the case of
$g_{\omega}$ its value is slightly higher in SINPA than in SINPB. The
pairing is treated within the BCS approximation with cut-off energy in
pairing space taken as $\hbar\omega_0 = 41A^{-1/3}$ MeV. The BCS pairing
strengths for neutron and proton for the models SINPB and SINPA were kept
fixed to $G_n=20/A$ and $G_p=25/A$.  The neutron and proton pairing gaps
($\Delta_n, \Delta_p$) in MeV for the neutron rich nuclei  are $^{30}$Ne
(0.0, 2.3), $^{36}$Mg (2.5, 2.0), $^{54}$Ca (1.1, 0.0), $^{58}$Ca (1.0,
0.0), $^{138}$Sn (1.3, 0.0). The pairing gaps for other non-magic nuclei
are close to $12/\sqrt{A}$ MeV. The neutron pairing gap for   $^{24}$O
practically vanishes, since, the first unoccupied $1d_{3/2}$ orbit is
about 4.5 MeV above the completely filled $2s_{1/2}$ orbit \cite{Chen15}.

In Table \ref{tab2} different observables $\mathcal{O}$ pertaining
to finite nuclei and neutron star, their experimental values, their
obtained values from SINPB and SINPA along with $\Delta\mathcal{O}$, the
adopted errors on them are listed. The experimental values of binding
energies of all the nuclei except for $^{54}$Ca used in the fit are
taken from the latest compilation AME-2012 \cite{Wang12}. Recently,
binding energy of $^{54}$Ca was measured very accurately at TRIUMF
\cite{Gallant12} and CERN \cite{Wienholtz13}.  For this present
calculation, the experimental value of the binding energy for $^{54}$Ca
is taken from Ref. \cite{Wienholtz13}.  Experimental values for the
charge radii used in the fit are obtained from the compilation by
Angeli and Marinova \cite{Angeli13}.  For the optimization of SINPA,
observed maximum mass of neutron star $M_{max}^{NS}$ is taken from
Ref. \cite{Demorest10, Antoniadis13}.  It may be pointed out that,
experimental value for some of the fit data are little different in the
present calculation in comparison to our previous paper \cite{Mondal15}.
Except for $^{68}$Ni, $\Delta\mathcal{O}$ for all the fit-data common to
both the models SINPB and SINPA are taken from Ref. \cite{Klupfel09}.
As the obtained value of binding energy of $^{68}$Ni from both the
models SINPB and SINPA deviate by more than 2 MeV from its experimental
value, demanding too much accuracy on that particular datum costs a
larger amount in total $\chi^2$ compared to other data points.  For this
reason we have taken $\Delta\mathcal{O} = 2$ MeV for the binding energy
of $^{68}$Ni unlike in Ref.\cite{Klupfel09}, where $\Delta\mathcal{O}
= 1$ MeV.  Calculated errors on the binding energies and charge radii
due to uncertainties in the model parameters for the fitted nuclei for
both the models SINPB and SINPA lie within the range from 0.51 - 1.89
MeV and 0.005 - 0.016 fm, respectively. In model SINPA the obtained
maximum neutron star mass $M_{max}^{NS}$ (1.98$\pm$0.03 $M_{\odot}$)
compares well with the observed value.  We would like to point out
that the two isotopes of Mg nuclei used in the optimization of SINPA
are deformed. The numerical computation is done with 20 oscillator
shells being taken as the basis states for the nucleons.  The quadrupole
deformation parameter $\beta_2$ calculated from SINPA for $^{24}$Mg and
$^{36}$Mg nuclei are found to be 0.47 and 0.37, respectively.

\subsection{Nuclear Matter Properties}
The energy per nucleon in asymmetric nuclear matter as a function of density 
$\rho$ and isospin asymmetry $\delta$ is approximately given by,
\begin{equation}
\mathcal{E} (\rho,\delta) = \mathcal{E} (\rho,0) + C_{sym} (\rho) 
\delta^2,
\label{eden}
\end{equation}
where, $\rho$ =($\rho_n+\rho_p$) and $\delta= \frac{\rho_n-\rho_p}{\rho}$.
The term $\mathcal{E} (\rho,0)$ represents the energy per nucleon
in symmetric nuclear matter and $C_{sym} (\rho)$ is the symmetry
energy. Energy per nucleon $\mathcal{E} (\rho,0)$ can be expressed in
terms of model parameters as,
\begin{eqnarray}
\label{erho}
\mathcal{E} (\rho,0)=&& \frac{2}{\pi^2}\int_0^{k_F} dk\ k^2 \sqrt{k^2+{M^*}^2}\nonumber
\\ &&+\frac{1}{2}m_{\sigma}^2 \sigma^2 + \frac{{\kappa_3}}{6M}
g_{\sigma}m_{\sigma}^2\sigma^3 + \frac{{\kappa_4}}{24M^2}g_{\sigma}^2
m_{\sigma}^2\sigma^4 \nonumber
\\ &&-\frac{1}{2}m_{\omega}^2 \omega^2 - \frac{1}{24}\zeta_0 g_{\omega}^{2} \omega^4,
\end{eqnarray}
and, $C_{sym}(\rho)$ is expressed as,
\begin{equation}
C_{sym}(\rho)=\frac{k_F^2}{6(k_F^2+{M^*}^2)^{1/2}}+\frac{g_{\boldsymbol
\rho}^2}{12\pi^2}\frac{k_F^3}{{m_{\boldsymbol\rho}^*}^2}.
\label{csymrho}
\end{equation}
Here, $k_F$ is the nucleon Fermi momentum in symmetric nuclear matter
at density $\rho$ (=$\frac{2k_F^3}{3\pi^2}$). The Dirac effective mass
of nucleon $M^*$ is given by,
\begin{equation}
M^*=M-g_{\sigma} \sigma,
\label{mstar}
\end{equation}
and, the effective mass of $\boldsymbol\rho$ meson,
$m_{\boldsymbol\rho}^*$ is expressed as \cite{Horowitz01a},
\begin{equation}
{m_{\boldsymbol\rho}^*}^2 = m_{\boldsymbol\rho}^2\left(1+\frac{1}{2M^2}
\eta_{2\boldsymbol\rho}g_{\omega}^2\omega^2\right).
\label{mrhostar}
\end{equation}
In the Eqs. (\ref{erho}-\ref{mrhostar}) the value of the fields
$\sigma$ and $\omega$ are obtained by solving their field equation
at a particular density $\rho$. From Eq. (\ref{csymrho}) one can see
that, the kinetic part of $C_{sym}(\rho)$ depends on the effective mass of
nucleon $M^*$, which has dependence on the parameter $g_{\sigma}$ and the
field value of $\sigma$. However, the interaction part of $C_{sym}(\rho)$
mainly depends on the isovector parameters $g_{\boldsymbol \rho}$ and
$\eta_{2\boldsymbol\rho}$.  The value of the incompressibility coefficient
$K_0$ at the saturation density is related to $\mathcal{E} (\rho,0)$ as,
\begin{equation}
K_0 =9\left[\rho^2\left(\frac{d^2\mathcal{E}(\rho,0)}{d\rho^2}\right)
\right]_{\rho=\rho_0}.
\label{K0}
\end{equation}
The symmetry energy slope parameter $L$ at a given density $\rho$ can be
evaluated as,
\begin{equation}
L = 3\rho \left(\frac{dC_{sym} (\rho)}{d\rho}\right).
\label{L}
\end{equation}

\begin{table}[t]
\caption{\label{tab3}
Different nuclear matter properties: the binding energy per nucleon for
symmetric matter $\mathcal E_0$, incompressibility coefficient $K_0$,
Dirac effective mass of nucleon $M^*_0$ (scaled by nucleon mass $M$),
symmetry energy coefficient $C_{sym}^0$ and density slope parameter of
symmetry energy $L_0$ for the nuclear matter evaluated at saturation
density $\rho_0$  along with the correlated errors on them for the models
SINPB and SINPA. The values of $C_{sym}(\rho_{c})$ and $L(\rho_{c})$
calculated at crossing density $\rho_c$ along with the neutron skin
$\Delta r_{np}$ in  $^{208}$Pb are also presented for these two models.  }
\begin{ruledtabular}
\begin{tabular}{ccc} Observable  & SINPB & SINPA\\ \hline 
$\mathcal E_0$ (MeV)           &  $-16.04\pm0.06$  & $-16.00\pm0.05$   \\ 
$K_0$ (MeV)             &  $206\pm20$   & $203\pm6$    \\
$\rho_0$ (fm$^{-3}$)  &  $0.150\pm0.002$    & $0.151\pm0.001$     \\
$M^*_0/M$               &  $0.59\pm0.01$    & $0.58\pm0.01$     \\
$C_{sym}^0$ (MeV)         &  $33.95\pm2.41$     & $31.20\pm1.11$      \\
$C_{sym}(\rho_{c})$ (MeV)         & $26.08\pm0.41$     & $25.60\pm0.51$     \\
$L_0$ (MeV)           &  $71.55\pm18.89$    & $53.86\pm4.66$      \\
$L(\rho_{c})$ (MeV)           &  $55.98\pm13.78$   & $38.47\pm5.43$     \\
$\Delta r_{np}$ ($^{208}$Pb) (fm) & $0.241\pm0.040$ & $0.183\pm0.022$\\
\end{tabular}
\end{ruledtabular}
\end{table}

Once the objective functions for the models SINPB and SINPA are
optimized, different nuclear matter properties can be extracted from
them and compared.  In Table \ref{tab3} values of different nuclear
matter parameters along with the corresponding errors evaluated within
the covariance analysis are listed for SINPB and SINPA.  The properties
associated with symmetric nuclear matter are evaluated at the saturation
density $\rho_0$, while, those characterizing the asymmetric nuclear
matter are evaluated at $\rho_0$ and  the crossing-density $\rho_c$
which is taken as $\frac{0.11}{0.16}\times\rho_0$ \cite{Wang15}.
Errors on binding energy per nucleon $\mathcal E_0$ ($=\mathcal{E}
(\rho_0,0)$), saturation density $\rho_0$ and Dirac effective mass of
nucleon $M^*_0/M$ (=$M^*(\rho_0)/M$) are pretty much the same for both
the models concerned.  However, a noticeable improvement is observed
for the model SINPA over SINPB for the calculated errors on the symmetry
energy parameters $C^0_{sym}$ (=$C_{sym}(\rho_0)$), $L_0$ (=$L(\rho_0)$)
and $L(\rho_c)$.  The refinement in the error in SINPA in comparison
to SINPB is also to be noted for the incompressibility coefficient at
saturation density, $K_0$.  Error on the neutron-skin $\Delta r_{np}$ in
$^{208}$Pb also reduces by almost a factor of $2$ in SINPA in comparison
to SINPB. The central values of $L_0$ and $\Delta r_{np}$ of $^{208}$Pb
obtained for the model SINPB are seen to differ from  those obtained
from the  model-I of Ref. \cite{Mondal15}; this can be attributed to the
differences in the adopted error on the binding energy of $^{68}$Ni and
to the differences in some of the experimental fit data.

The observation of improved constraint in the symmetry elements
calculated from model SINPA over those from SINPB  clearly indicates
that the additional data of four highly asymmetric nuclei ($^{24}$O,
$^{30}$Ne, $^{36}$Mg and $^{58}$Ca) with $\delta>0.3$ and the observed
maximum mass of neutron star $M_{max}^{NS}$ contain more distilled
information on isovector elements in the nuclear interaction.  It is
striking to note that the addition of the binding energies of $^{54}$Ca,
$^{78}$Ni and $^{138}$Sn ($\delta \sim 0.26$-$0.28$) as fit data in the
optimization of the model SINPB did  not improve the uncertainties in
the  symmetry energy parameters as compared to those for the model-I in
Ref. \cite{Mondal15}. On the other hand, inclusion of  highly asymmetric
($\delta >0.3$) $^{36}$Mg and $^{58}$Ca nuclei in the fitting protocol of
the model SINPA yields smaller uncertainties in the symmetry energy
parameters in comparison to the model-II of Ref. \cite{Mondal15} which
does not include these nuclei.  This clearly emphasizes that the binding
energies of nuclei with $\delta > 0.3$ play a crucial role in constraining
the symmetry energy parameters and is thus a pointer to the necessity
of taking data for very asymmetric nuclei in the optimization of the RMF
model. In the next section we are going to analyze this more critically.
\begin{figure}[h]{}
{\includegraphics[height=3.5in,width=3.2in,angle=-90]{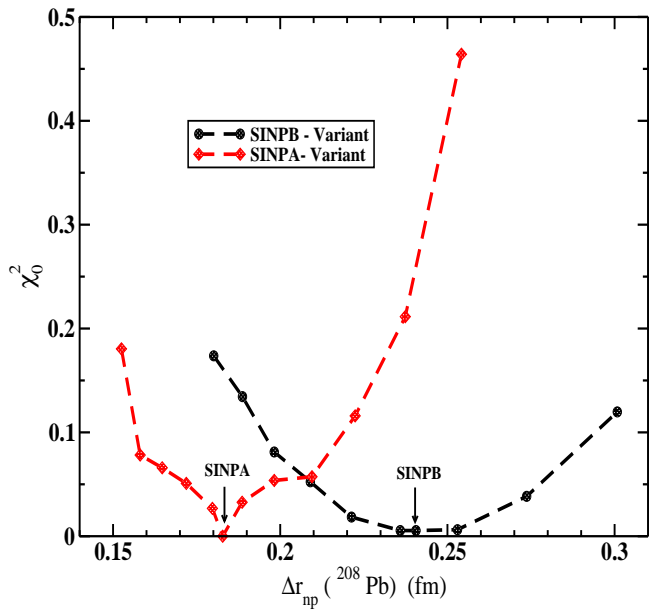}}
\caption{\label{fig1}
(Color online) Optimum values of the objective function ($\chi_0^2$)
are plotted as a function of $\Delta r_{np}$ (neutron skin of $^{208}$Pb)
for two families of models, namely, SINPB-Variant and SINPA-Variant
(see text for details).}
\end{figure}

\subsection{Sensitivity analysis }
We now look into the sensitivity of symmetry energy parameters to the
properties of the neutron rich systems in some detail.  Before embarking
on our analysis in terms of sensitivity matrix, we make a simple
examination of the results.  We look into the dependence of the optimal
value of the objective function   on the neutron skin of $^{208}$Pb.
Fixing $\eta_{2\boldsymbol\rho}$ to a preset value  and optimizing the
$\chi^2$ function by adjusting the rest of the model parameters,  one
can get a particular value of $\Delta r_{np}$ of $^{208}$Pb for the
models SINPB and SINPA \cite{Sil05}.  Two families of RMF models so
constructed are called SINPB-Variant and SINPA-Variant. Different input
values of $\eta_{2\boldsymbol\rho}$ would yield different $\Delta r_{np}$
in both these models. In Fig. \ref{fig1} optimal values of the objective
function $\chi^2$ (i.e. $\chi_0^2$) for these two models are displayed
as a function of $\Delta r_{np}$ of $^{208}$Pb; the values of $\chi_0^2$
are so adjusted that their minimum value within a family vanishes. Visual
comparison of results from the two families of models shows that there
is a stronger
\begin{figure}[h]{}
{\includegraphics[height=3.5in,width=3.2in,angle=-90]{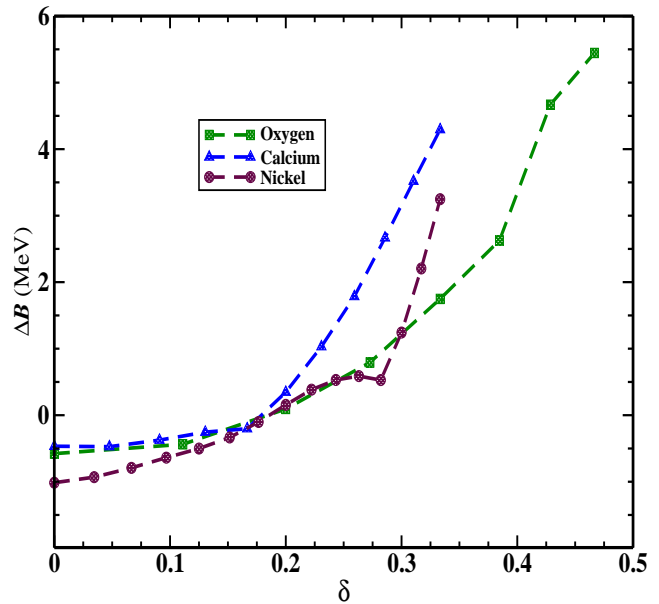}}
\caption{\label{fig2}
(Color online) Binding energy differences $\Delta B$ (=$B({\rm
SINPB})-B({\rm SINPA})$) extracted using models SINPB and SINPA for
even isotopes of O, Ca and Ni nuclei plotted as a function of asymmetry
$\delta$.}
\end{figure}
preference to a particular value of $\Delta r_{np}$ of $^{208}$Pb in the
SINPA-Variant family.  It is worthwhile to mention that, SINPB-Variant
family has $^{54}$Ca, $^{78}$Ni and $^{138}$Sn in the fitted data set
where asymmetry $\delta \sim 0.26$ - $0.28$. The $\chi_0^2$ function is
still rather flat, making it tenuous to give a reasonable bound on the
value of $\Delta r_{np}$ of $^{208}$Pb. The role of ultra neutron-rich
nuclei in the SINPA-Variant family where nuclei with $\delta > 0.3$
(e.g. $^{24}$O, $^{30}$Ne, $^{36}$Mg, $^{58}$Ca) are further included in
the fitting protocol are eminently evident in Fig \ref{fig1}.  As $\Delta
r_{np}$ of $^{208}$Pb is correlated to $L_0$ \cite{Centelles09,Chen05},
one finds a tighter constraint on $L_0$ as well from SINPA as compared
to SINPB (see Tab. \ref{tab3}).

The two Variant families so constructed from selective optimization
of the parameter set ${\bf p_0}$ keeping $\Delta r_{np}$ of $^{208}$Pb
fixed should affect the calculated binding energies. In Fig. \ref{fig2}
binding energy differences of three isotopic chains of O, Ca and Ni
extracted from models SINPB  and SINPA ($\Delta r_{np}$ ($^{208}$Pb) =
0.241 fm and 0.183 fm, respectively at absolute minima of $\chi_0^2$,
see Fig. \ref{fig1}) are plotted as a function of asymmetry $\delta$.
The differences in the binding energies so calculated for all the isotopic
chains show significant enhancement when one goes from $\delta$ just below
0.3 to higher values \cite{Chen15}. Nuclei beyond $\delta$ = 0.3 thus
show a high sensitivity towards $\Delta r_{np}$ of $^{208}$Pb. Several
experimental efforts are being made to accurately measure binding
energies of these exotic nuclei.  These measurements may impose very
tight constraint on the value of $\Delta r_{np}$ of $^{208}$Pb.

The sensitivity of a given parameter to a particular data can be
determined in terms of the sensitivity matrix of dimension $N_p \times
N_d$ defined as \cite{Kortelainen10,Dobaczewski14},
\begin{equation}
S({\bf p})= [\hat J({\bf p}) \hat J^T({\bf p})]^{-1}\hat J({\bf p}).
\label{smat}
\end{equation}
Here $\hat J({\bf p})$ is the Jacobian matrix with the same dimension
as $S({\bf p})$; its elements are given as,
\begin{equation}
\hat J_{\alpha i}= \frac{1}{\Delta \mathcal{O}_i}\left(\frac{\partial 
\mathcal{O}_i}{\partial p_{\alpha}}\right)_{\bf p_0}.
\label{jmat}
\end{equation}
The sensitivity of the parameter $p_{\alpha}$ to the i-th data is given
by $S_{\alpha i}^2$ which is normalized to $\sum_{i=1}^{N_d} S_{\alpha
i}^2 = 1$. The relative sensitivity for a subset of data can likewise
be obtained by summing $S_{\alpha i}^2$ over that subset.
\begin{figure}[h]{}
{\includegraphics[height=3.3in,width=3.5in]{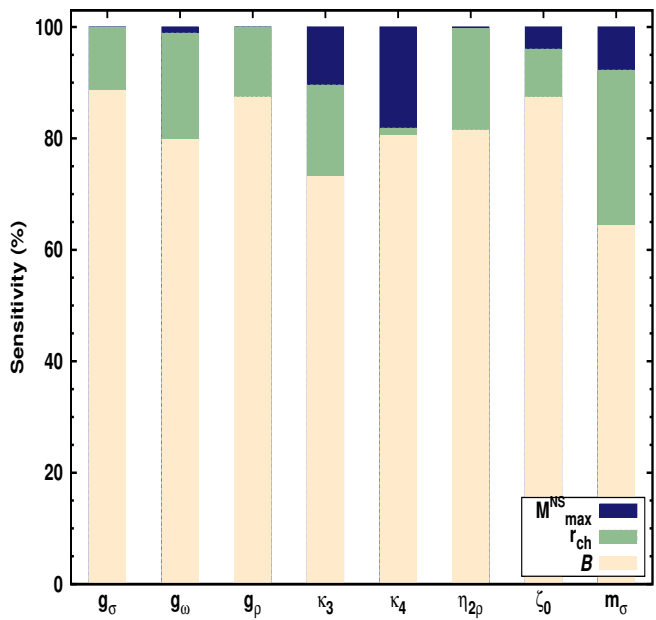}}
\caption{\label{fig3}
(Color online) Relative sensitivity of different parameters of the
effective Lagrangian density to three groups of fit data used in
optimization of SINPA. These groups are nuclear binding energies
($B$), charge radii ($r_{ch}$) and maximum mass of neutron star
($M_{max}^{NS}$).}
\end{figure}

\begin{figure}[h]{}
{\includegraphics[height=3.3in,width=3.5in]{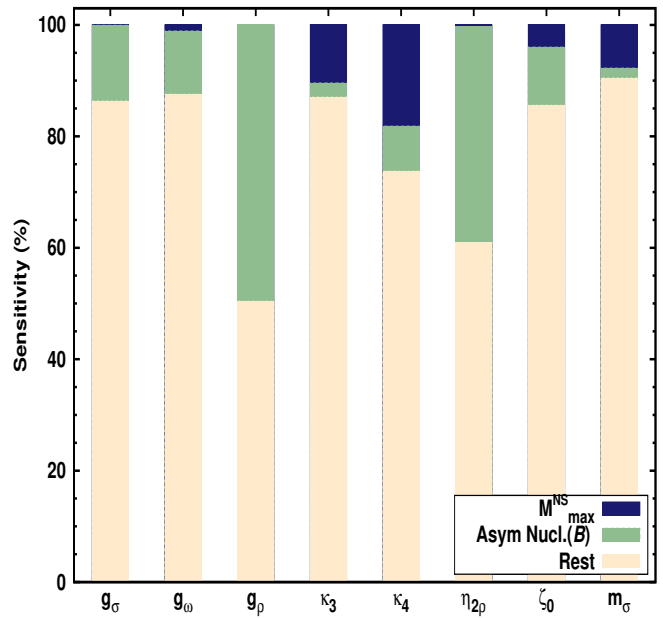}}
\caption{\label{fig4}
(Color online)  Same as Fig. \ref{fig3}, but, with different grouping
of the fit data of finite nuclei. One group contains binding energies of
highly asymmetric nuclei ($^{24}$O, $^{30}$Ne, $^{36}$Mg and $^{58}$Ca)
and another contains remaining fit data on the finite nuclei.}
\end{figure}
We employed the sensitivity analysis in model SINPA to understand the
impact of the new fit-data considered to optimize it.  In Fig. \ref{fig3}
the relative sensitivity of different parameters of the effective
Lagrangian density to three broad data-types (binding energies $B$,
charge radii $r_{ch}$ of finite nuclei and maximum mass of neutron star
$M_{max}^{NS}$) are displayed.  It is evident that all the parameters
 are maximally sensitive ($>$65$\%$) to the binding energies of nuclei.
The higher relative sensitivity of the parameters to the binding energies
of nuclei can be attributed partly to their large number used in the
fit.  The parameter $\kappa_4$ shows almost no sensitivity towards the
charge radii.  The parameters $\kappa_3$, $\kappa_4$ and $m_{\sigma}$
are seen to be appreciably sensitive to the single data of neutron star
$M_{max}^{NS}$ as they have a crucial role in the determination of the
high density behavior of the nuclear EoS which in turn governs the value
of $M_{max}^{NS}$.

In Fig. \ref{fig4} we perform the analysis by regrouping the data on
binding energies and charge radii so that the sensitivity of the RMF
model parameters to the binding energies of highly asymmetric nuclei can
be assessed.  One of the group consists of only the binding energies of
$^{24}$O, $^{30}$Ne, $^{36}$Mg and $^{58}$Ca nuclei, while the other
group contains the remaining data on the finite nuclei.  One can not
fail to notice that, the parameters $g_{\mathbf{\boldsymbol\rho}}$ and
${\eta_{2\boldsymbol\rho}}$, which control the isovector part of the
effective Lagrangian, are relatively more sensitive ($\sim 40\%$) to
the binding energies of highly asymmetric nuclei.  The sensitivity
of $g_{\boldsymbol\rho}$ and ${\eta_{2\boldsymbol\rho}}$ to the value
of $M_{max}^{NS}$ is not observed in Figs. \ref{fig3} and \ref{fig4}
partly because $M_{max}^{NS}$ is a single datum, but mainly because it
is overshadowed by the relative contributions to the sensitivity from
the binding energies of asymmetric nuclei.

\begin{figure}[t]{}
{\includegraphics[height=3.3in,width=3.5in]{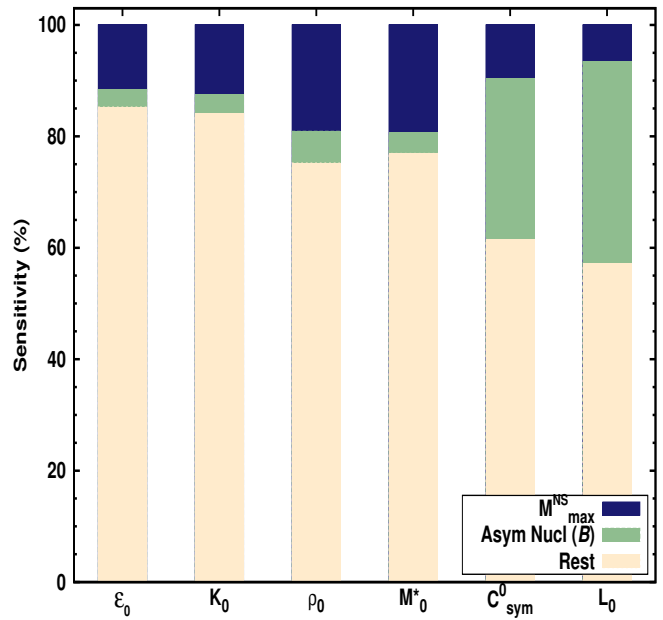}}
\caption{\label{fig5}
(Color online) Relative sensitivity of the nuclear matter properties at
saturation density to the fit data of SINPA with the same grouping as
in Fig. \ref{fig4}.}
\end{figure}

In Fig. \ref{fig5} we display at saturation density $\rho_0$ the
sensitivity of different empirical data pertaining to the nuclear matter
to the data-set used in the optimization of the model SINPA. To do so,
we used the same grouping of data as in Fig. \ref{fig4}. Since the
parameters $g_{\sigma}$, $g_{\omega}$ etc. of the effective Lagrangian
are optimally determined from the full data set, it is no wonder that the
empirical nuclear matter data obtained from the energy density functional
are maximally sensitive to the group of fit data "Rest", as it contains
the largest number of data elements.  The high sensitivity of $C_{sym}^0$
($\sim 30\%$) and $L_0$ ($\sim 40\%$) to the binding energies of the
highly asymmetric $^{24}$O, $^{30}$Ne, $^{36}$Mg and $^{58}$Ca nuclei,
which form a very small subset of the data-set used in the optimization
of SINPA (4 out of 30) is a reflection of the high sensitivity of the
model parameters $g_{\boldsymbol\rho}$ and ${\eta_{2\boldsymbol\rho}}$
to the masses of these highly asymmetric nuclei as seen earlier in
Fig. \ref{fig4}.  Appreciable sensitivity of all the nuclear matter
properties to the single data on neutron star $M_{max}^{NS}$ can not
also be missed either.  Accurate knowledge of $M_{max}^{NS}$ is required
for the precision determination of the EDF involving high densities
beyond saturation, any small change in it thus may result in large
change in the value of the nuclear matter properties ($\mathcal E_0,
K_0, \rho_0, M^*_0$) calculated from the EDF. This can be appreciated
from the sensitivity of $\kappa_3, \kappa_4$ and partly $\zeta_0$
(governing the scalar mass and the number density) on $M_{max}^{NS}$
displayed in Fig. \ref{fig4}.  The not-too-insignificant sensitivity of
$C_{sym}^0$ and $L_0$ to $M_{max}^{NS}$ demands attention.  It stems
from the dependence of the kinetic part of $C_{sym}(\rho)$ on $M^*$
(Eq. (\ref{csymrho})) whose value at saturation density is found
appreciably sensitive to the maximum mass of neutron star.  The value
of $\sigma$-field  determining the effective mass of nucleon depends
on the coupling constants $g_{\sigma}$, $\kappa_3$, $\kappa_4$ and the
value of $m_{\sigma}$.  High sensitivity of these coupling constants to
$M_{max}^{NS}$ (see Figs. \ref{fig3} and \ref{fig4}) gets reflected in the
sensitivity analysis of the symmetry energy parameters to $M_{max}^{NS}$.

To this end, we would like to mention that, the binding energies for
finite nuclei near neutron drip line as considered in the present work may
be  in general sensitive to the way pairing correlations are treated. We
compare our results with those obtained in Ref.  \cite{Chen15} calculated
for neutron-rich nuclei for the same form of Lagrangian density but with
'exact' treatment of pairing. We find that the sensitivity of binding
energies of neutron-rich  isotopes of Oxygen and Calcium nuclei to the
$\Delta r_{\rm np}$ of $^{208}$Pb is very similar.  To be specific, in
Ref. \cite{Chen15}, the binding energy of $^{24}$O increases by $\sim 4$
MeV when the neutron-skin thickness ($\Delta r_{\rm np}$) of $^{208}$Pb
changes from 0.16 fm to 0.28 fm. Similar trend is observed in our present
calculation. For $^{24}$O, the gain in binding energy is $\sim 2$ MeV when
$\Delta r_{\rm np}$ changes from 0.18 fm (SINPA) to 0.24 fm (SINPB). The
same feature is seen for the case of $^{58}$Ca. However, for more precise
determination of neutron-skin thickness in heavy nuclei on the basis of
nuclear masses, Relativistic Hartree Bogoliubov (RHB) calculation for
pairing for nuclei near the drip line is preferable \cite{Vretenar03,
Niksic14, Lalazissis99a}; this was not pursued in the present calculation.

\subsection{Nuclear Matter properties at high density}
\begin{figure}[h]{}
{\includegraphics[height=4.5in,width=3.5in]{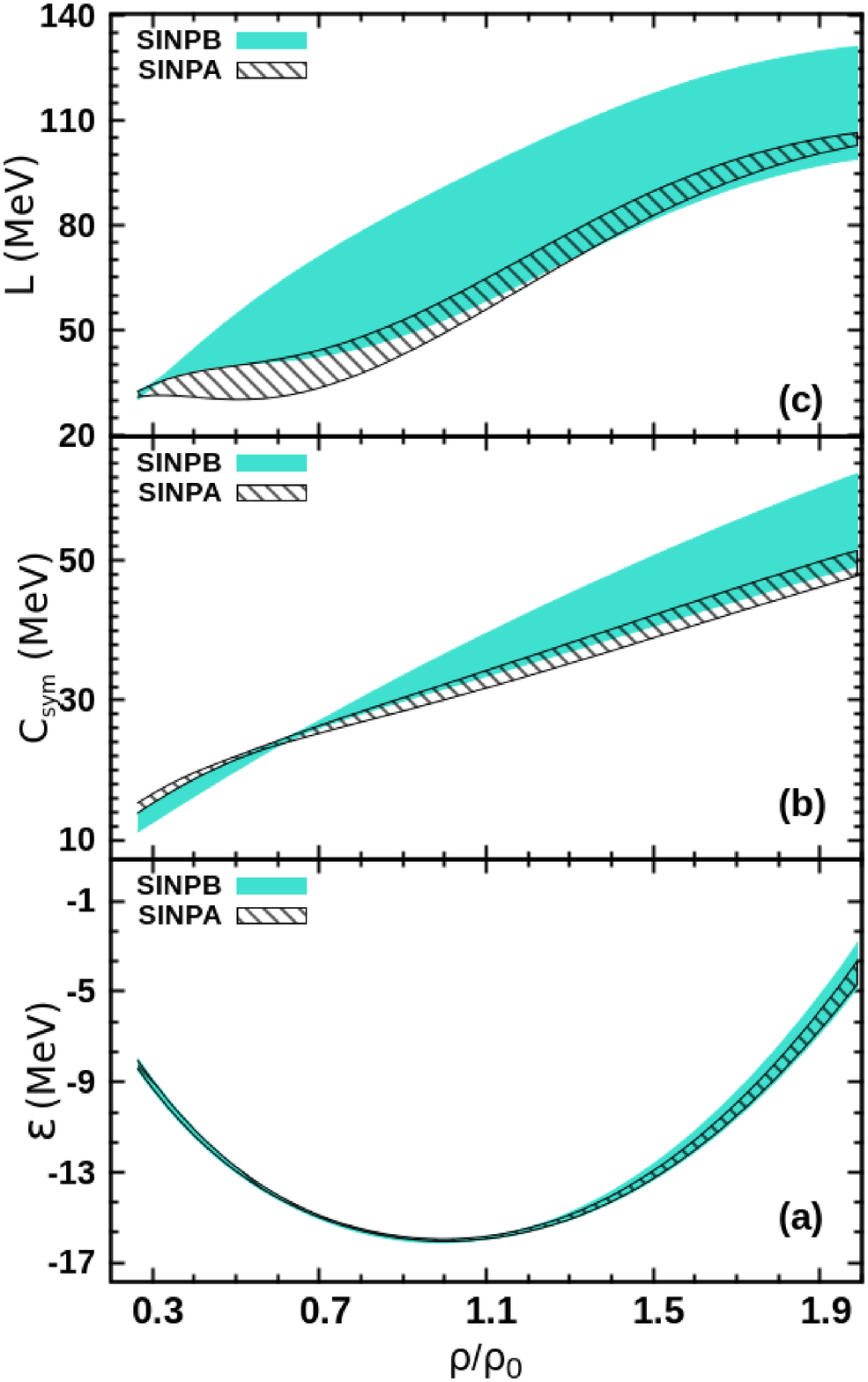}}
\caption{\label{fig6}
(Color online) Binding energy per nucleon for symmetric matter $\mathcal E$,
symmetry energy parameter $C_{sym}$ and its density derivative $L$ along
with their errors as a function of density $\rho/\rho_0$ for SINPB and
SINPA. }
\end{figure}
We extended our calculation of nuclear matter properties with both the
models SINPB and SINPA at densities beyond saturation.  This provides
valuable informations to construct theories for dense nuclear systems
viz. neutron star and several other astrophysical objects from EoS
so constrained at saturation density. In Fig. \ref{fig6} we have
plotted different nuclear matter properties, e.g. binding energy
per nucleon for symmetric matter $\mathcal E$ (Fig. \ref{fig6}(a)),
symmetry energy coefficient $C_{sym}$ (Fig \ref {fig6}(b)) and its
density derivative $L$ (Fig. \ref{fig6}(c)) as a function of density
$\rho/\rho_0$ for the models SINPB (turquoise) and SINPA (black-pattern)
along with their associated errors.  The errors are calculated within
the covariance analysis.  The energy per nucleon $\mathcal E$ in the
explored density region for SINPB and SINPA are almost identical as
seen from Fig. \ref{fig6}(a). Most stringent constraint on the values of
$\mathcal E$ appear at $\rho \sim \rho_0$ for both the models and they
grow as one moves away from $\rho_0$ \cite{De15}. In Fig.  \ref{fig6}(b)
allowed regions of $C_{sym}$ show similar trend for SINPB and SINPA,
both of them having their minimum variance at $\rho \sim 0.7\rho_0$
\cite{Zhang13}. However, a significant improvement is observed over
the errors on $C_{sym}$ for SINPA in comparison to SINPB at higher
densities.  Comparison of calculated electric dipole polarizability of
$^{208}$Pb from several Skyrme and RMF interactions with the corresponding
experimental data recently yielded a very tightly constrained value of
$C_{sym}$ at density $\rho_0/3$, $C_{sym} (\rho_0/3)= 15.91\pm0.99$ MeV
\cite{Zhang15}. It is interesting to note that the model SINPB has overlap
with this constraint at the lower end, $C_{sym} (\rho_0/3) = 13.69$ -
$16.31$ MeV, whereas SINPA agrees with this result at the higher end,
$C_{sym} (\rho_0/3)= 16.41$ - $17.67$ MeV.

\begin{figure}[t]{}
{\includegraphics[height=3.5in,width=3.2in,angle=-90]{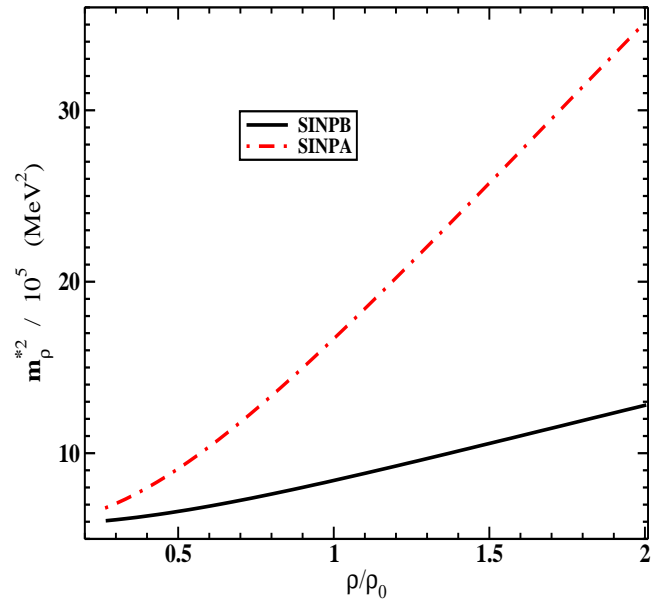}}
\caption{\label{fig7}
(Color online) Square of the effective mass of $\boldsymbol\rho$ meson
(scaled by $10^5$) as a function of density $\rho/\rho_0$ plotted for
SINPB and SINPA. }
\end{figure}
In Fig. \ref{fig6}(c) a curious behavior in the variance of $L$ with
density was observed.  For the model SINPB, the variance in $L$ grows
up to a certain density $\sim \rho_0$ and from there onwards it remained
almost constant all the way up to $2\rho_0$. In contrast, in SINPA error
on $L$ grows only up to $\rho \sim 0.7\rho_0$ and shows a monotonically
decreasing trend afterwards.  This particular result may appear
intriguing.  A model primarily obtained by fitting some ground state
properties of finite nuclei, where concerned central density is $\sim
\rho_0$ and average density is $\sim 0.7\rho_0$ is not normally expected
to show better constraint on nuclear matter properties at ultra-saturation
densities. To investigate this, we looked into the expression of $C_{sym}$
as a function of density given in Eq. (\ref{csymrho}). $C_{sym}$
has a dependence on ${m_{\boldsymbol\rho}^*}^2$, the square of
the effective mass of ${\boldsymbol\rho}$ meson.  The density
variation of ${m_{\boldsymbol\rho}^*}^2$ for both the models are
displayed in Fig. \ref{fig7}.  A rapid difference in the value of
${m_{\boldsymbol\rho}^*}^2$ (scaled by $10^5$) calculated in models SINPA
and SINPB builds up with increasing density. As the value of the parameter
$\eta_{2\boldsymbol\rho}$ is much larger in SINPA (38.18) compared to
that in SINPB (13.49) [see Table \ref{tab1}], at high densities the
second term in the expression of $C_{sym}$ (Eq. (\ref{csymrho})) gets
diluted due to ${m_{\boldsymbol\rho}^*} ^2$ (Eq. (\ref{mrhostar}))
by a much greater rate for the model SINPA in comparison to SINPB.
This explains why the error on $C_{sym}$ grows at much faster rate in
SINPB than in SINPA. Now, if one takes density derivative of $C_{sym}$,
the second term in Eq. (\ref{csymrho}) gives rise to two terms with
$\eta_{2 \boldsymbol\rho}$ in the denominator for the expression of $L$
as a function of density due to varying $\omega$ field value.  That is why
$\eta_{2\boldsymbol\rho}$ becomes a very crucial factor for the values
of $L$ at higher densities. This fact explains why in SINPA error on
$L$ decreases at higher densities, whereas in SINPB it remains almost
constant as shown in Fig. \ref{fig6}(c).


\section {Summary and Conclusions}
To sum up, we have made an investigation in this paper on
the extraction of the precision information from experimental data on
the isovector content of the nuclear interaction and their observable
derivatives like the symmetry energy of nuclear matter and its density
slope $L_0$ at saturation density. The relativistic mean field model is
chosen as the framework for the realization of this goal. A comparative
study of the covariance analysis of the interaction strengths and the
symmetry observables ($C_{sym}^0$, $L_0$, $\Delta r_{np}$ of $^{208}$Pb)
made with two models SINPB (one that included in the fit data selective
isovector sensitive information on observables from nearly symmetric and
asymmetric nuclei) and SINPA (which included further data from extremely
asymmetric nuclei right at the edge of neutron drip line with neutron
to proton ratio $\sim$ 2 and the observed maximum mass $M_{max}^{NS}$
of neutron star) shows that the nuclear symmetry energy properties and
the neutron skin $\Delta r_{np}$ of $^{208}$Pb are determined in much
narrower constraints from the latter model.  This is a pointer to the
necessity of inclusion of extremely neutron-rich systems in any data
analysis for filtering out information on isovector entities in the
nuclear interaction. Noticeably growing difference in the binding energies
of isotopes of some chosen nuclei with nuclear asymmetry $\delta$ beyond
$\sim 0.3$, calculated in models SINPA and SINPB and also in a somewhat
related paper \cite{Chen15} tend to confirm this.  The conclusion
is further reinforced  from the sensitivity analysis of the
different model parameters entering the nuclear effective interaction
to the experimental data set taken for such an analysis. To check the
robustness  of such calculational outcome, it may be interesting to extend
the investigation to other different versions of the mean-field models,
both relativistic and non-relativistic.


\begin{thebibliography}{48}
\expandafter\ifx\csname natexlab\endcsname\relax\def\natexlab#1{#1}\fi
\expandafter\ifx\csname bibnamefont\endcsname\relax
  \def\bibnamefont#1{#1}\fi
\expandafter\ifx\csname bibfnamefont\endcsname\relax
  \def\bibfnamefont#1{#1}\fi
\expandafter\ifx\csname citenamefont\endcsname\relax
  \def\citenamefont#1{#1}\fi
\expandafter\ifx\csname url\endcsname\relax
  \def\url#1{\texttt{#1}}\fi
\expandafter\ifx\csname urlprefix\endcsname\relax\def\urlprefix{URL }\fi
\providecommand{\bibinfo}[2]{#2}
\providecommand{\eprint}[2][]{\url{#2}}

\bibitem[{\citenamefont{Centelles et~al.}(2009)\citenamefont{Centelles,
  Roca-Maza, Vi{\~n}as, and Warda}}]{Centelles09}
\bibinfo{author}{\bibfnamefont{M.}~\bibnamefont{Centelles}},
  \bibinfo{author}{\bibfnamefont{X.}~\bibnamefont{Roca-Maza}},
  \bibinfo{author}{\bibfnamefont{X.}~\bibnamefont{Vi{\~n}as}},
  \bibnamefont{and} \bibinfo{author}{\bibfnamefont{M.}~\bibnamefont{Warda}},
  \bibinfo{journal}{Phys. Rev. Lett.} \textbf{\bibinfo{volume}{102}},
  \bibinfo{pages}{122502} (\bibinfo{year}{2009}).

\bibitem[{\citenamefont{M{\"o}ller et~al.}(2012)\citenamefont{M{\"o}ller,
  Myers, Sagawa, and Yoshida}}]{Moller12}
\bibinfo{author}{\bibfnamefont{P.}~\bibnamefont{M{\"o}ller}},
  \bibinfo{author}{\bibfnamefont{W.~D.} \bibnamefont{Myers}},
  \bibinfo{author}{\bibfnamefont{H.}~\bibnamefont{Sagawa}}, \bibnamefont{and}
  \bibinfo{author}{\bibfnamefont{S.}~\bibnamefont{Yoshida}},
  \bibinfo{journal}{Phys. Rev. Lett} \textbf{\bibinfo{volume}{108}},
  \bibinfo{pages}{052501} (\bibinfo{year}{2012}).

\bibitem[{\citenamefont{Jiang et~al.}(2012)\citenamefont{Jiang, Fu, Zhao, and
  Arima}}]{Jiang12}
\bibinfo{author}{\bibfnamefont{H.}~\bibnamefont{Jiang}},
  \bibinfo{author}{\bibfnamefont{G.~J.} \bibnamefont{Fu}},
  \bibinfo{author}{\bibfnamefont{Y.~M.} \bibnamefont{Zhao}}, \bibnamefont{and}
  \bibinfo{author}{\bibfnamefont{A.}~\bibnamefont{Arima}},
  \bibinfo{journal}{Phys. Rev. C} \textbf{\bibinfo{volume}{85}},
  \bibinfo{pages}{024301} (\bibinfo{year}{2012}).

\bibitem[{\citenamefont{Agrawal et~al.}(2012)\citenamefont{Agrawal, De, and
  Samaddar}}]{Agrawal12}
\bibinfo{author}{\bibfnamefont{B.~K.} \bibnamefont{Agrawal}},
  \bibinfo{author}{\bibfnamefont{J.~N.} \bibnamefont{De}}, \bibnamefont{and}
  \bibinfo{author}{\bibfnamefont{S.~K.} \bibnamefont{Samaddar}},
  \bibinfo{journal}{Phys. Rev. Lett.} \textbf{\bibinfo{volume}{109}},
  \bibinfo{pages}{262501} (\bibinfo{year}{2012}).

\bibitem[{\citenamefont{Agrawal et~al.}(2013)\citenamefont{Agrawal, De,
  Samaddar, Col{\`o}, and Sulaksono}}]{Agrawal13}
\bibinfo{author}{\bibfnamefont{B.~K.} \bibnamefont{Agrawal}},
  \bibinfo{author}{\bibfnamefont{J.~N.} \bibnamefont{De}},
  \bibinfo{author}{\bibfnamefont{S.~K.} \bibnamefont{Samaddar}},
  \bibinfo{author}{\bibfnamefont{G.}~\bibnamefont{Col{\`o}}}, \bibnamefont{and}
  \bibinfo{author}{\bibfnamefont{A.}~\bibnamefont{Sulaksono}},
  \bibinfo{journal}{Phys. Rev. C} \textbf{\bibinfo{volume}{87}},
  \bibinfo{pages}{051306(R)} (\bibinfo{year}{2013}).

\bibitem[{\citenamefont{Liu et~al.}(2013)\citenamefont{Liu, Ren, Xu, and
  Xu}}]{Liu13}
\bibinfo{author}{\bibfnamefont{J.}~\bibnamefont{Liu}},
  \bibinfo{author}{\bibfnamefont{Z.}~\bibnamefont{Ren}},
  \bibinfo{author}{\bibfnamefont{C.}~\bibnamefont{Xu}}, \bibnamefont{and}
  \bibinfo{author}{\bibfnamefont{R.}~\bibnamefont{Xu}}, \bibinfo{journal}{Phys.
  Rev. C} \textbf{\bibinfo{volume}{88}}, \bibinfo{pages}{024324}
  (\bibinfo{year}{2013}).

\bibitem[{\citenamefont{Danielewicz and Lee}(2009)}]{Danielewicz09}
\bibinfo{author}{\bibfnamefont{P.}~\bibnamefont{Danielewicz}} \bibnamefont{and}
  \bibinfo{author}{\bibfnamefont{J.}~\bibnamefont{Lee}},
  \bibinfo{journal}{Nuclear Physics A} \textbf{\bibinfo{volume}{818}},
  \bibinfo{pages}{36 } (\bibinfo{year}{2009}), ISSN \bibinfo{issn}{0375-9474}.

\bibitem[{\citenamefont{Liu et~al.}(2010)\citenamefont{Liu, Wang, Li, and
  Zhang}}]{Liu10}
\bibinfo{author}{\bibfnamefont{M.}~\bibnamefont{Liu}},
  \bibinfo{author}{\bibfnamefont{N.}~\bibnamefont{Wang}},
  \bibinfo{author}{\bibfnamefont{Z.}~\bibnamefont{Li}}, \bibnamefont{and}
  \bibinfo{author}{\bibfnamefont{F.}~\bibnamefont{Zhang}},
  \bibinfo{journal}{Phys. Rev. C} \textbf{\bibinfo{volume}{82}},
  \bibinfo{pages}{064306} (\bibinfo{year}{2010}).

\bibitem[{\citenamefont{Wang et~al.}(2013)\citenamefont{Wang, Ou, and
  Liu}}]{Wang13}
\bibinfo{author}{\bibfnamefont{N.}~\bibnamefont{Wang}},
  \bibinfo{author}{\bibfnamefont{L.}~\bibnamefont{Ou}}, \bibnamefont{and}
  \bibinfo{author}{\bibfnamefont{M.}~\bibnamefont{Liu}},
  \bibinfo{journal}{Phys. Rev. C} \textbf{\bibinfo{volume}{87}},
  \bibinfo{pages}{034327} (\bibinfo{year}{2013}).

\bibitem[{\citenamefont{Shetty et~al.}(2007)\citenamefont{Shetty, Yennello, and
  Souliotis}}]{Shetty07}
\bibinfo{author}{\bibfnamefont{D.~V.} \bibnamefont{Shetty}},
  \bibinfo{author}{\bibfnamefont{S.~J.} \bibnamefont{Yennello}},
  \bibnamefont{and} \bibinfo{author}{\bibfnamefont{G.~A.}
  \bibnamefont{Souliotis}}, \bibinfo{journal}{Phys. Rev. C}
  \textbf{\bibinfo{volume}{75}}, \bibinfo{pages}{034602}
  (\bibinfo{year}{2007}).

\bibitem[{\citenamefont{Famiano and {\it et. al}}(2006)}]{Famiano06}
\bibinfo{author}{\bibfnamefont{M.~A.} \bibnamefont{Famiano}} \bibnamefont{and}
  \bibinfo{author}{\bibnamefont{{\it et. al}}}, \bibinfo{journal}{Phys. Rev.
  Lett.} \textbf{\bibinfo{volume}{97}}, \bibinfo{pages}{052701}
  (\bibinfo{year}{2006}).

\bibitem[{\citenamefont{Li et~al.}(2008)\citenamefont{Li, Chen, and Ko}}]{Li08}
\bibinfo{author}{\bibfnamefont{B.-A.} \bibnamefont{Li}},
  \bibinfo{author}{\bibfnamefont{L.-W.} \bibnamefont{Chen}}, \bibnamefont{and}
  \bibinfo{author}{\bibfnamefont{C.~M.} \bibnamefont{Ko}},
  \bibinfo{journal}{Phys. Rep.} \textbf{\bibinfo{volume}{464}},
  \bibinfo{pages}{113} (\bibinfo{year}{2008}).

\bibitem[{\citenamefont{Roca-Maza
  et~al.}(2013{\natexlab{a}})\citenamefont{Roca-Maza, Brenna, Agrawal,
  Bortignon, Col{\`o}, Cao, Paar, and Vretenar}}]{Roca-Maza13}
\bibinfo{author}{\bibfnamefont{X.}~\bibnamefont{Roca-Maza}},
  \bibinfo{author}{\bibfnamefont{M.}~\bibnamefont{Brenna}},
  \bibinfo{author}{\bibfnamefont{B.~K.} \bibnamefont{Agrawal}},
  \bibinfo{author}{\bibfnamefont{P.~F.} \bibnamefont{Bortignon}},
  \bibinfo{author}{\bibfnamefont{G.}~\bibnamefont{Col{\`o}}},
  \bibinfo{author}{\bibfnamefont{L.-G.} \bibnamefont{Cao}},
  \bibinfo{author}{\bibfnamefont{N.}~\bibnamefont{Paar}}, \bibnamefont{and}
  \bibinfo{author}{\bibfnamefont{D.}~\bibnamefont{Vretenar}},
  \bibinfo{journal}{Phys. Rev. C} \textbf{\bibinfo{volume}{87}},
  \bibinfo{pages}{034301} (\bibinfo{year}{2013}{\natexlab{a}}).

\bibitem[{\citenamefont{Roca-Maza
  et~al.}(2013{\natexlab{b}})\citenamefont{Roca-Maza, Brenna, Col\`o,
  Centelles, Vi\~nas, Agrawal, Paar, Vretenar, and Piekarewicz}}]{Roca-Maza13a}
\bibinfo{author}{\bibfnamefont{X.}~\bibnamefont{Roca-Maza}},
  \bibinfo{author}{\bibfnamefont{M.}~\bibnamefont{Brenna}},
  \bibinfo{author}{\bibfnamefont{G.}~\bibnamefont{Col\`o}},
  \bibinfo{author}{\bibfnamefont{M.}~\bibnamefont{Centelles}},
  \bibinfo{author}{\bibfnamefont{X.}~\bibnamefont{Vi\~nas}},
  \bibinfo{author}{\bibfnamefont{B.~K.} \bibnamefont{Agrawal}},
  \bibinfo{author}{\bibfnamefont{N.}~\bibnamefont{Paar}},
  \bibinfo{author}{\bibfnamefont{D.}~\bibnamefont{Vretenar}}, \bibnamefont{and}
  \bibinfo{author}{\bibfnamefont{J.}~\bibnamefont{Piekarewicz}},
  \bibinfo{journal}{Phys. Rev. C} \textbf{\bibinfo{volume}{88}},
  \bibinfo{pages}{024316} (\bibinfo{year}{2013}{\natexlab{b}}).

\bibitem[{\citenamefont{Steiner and Gandolfi}(2012)}]{Steiner12}
\bibinfo{author}{\bibfnamefont{A.~W.} \bibnamefont{Steiner}} \bibnamefont{and}
  \bibinfo{author}{\bibfnamefont{S.}~\bibnamefont{Gandolfi}},
  \bibinfo{journal}{Phys. Rev. Lett.} \textbf{\bibinfo{volume}{108}},
  \bibinfo{pages}{081102} (\bibinfo{year}{2012}).

\bibitem[{\citenamefont{Dutra et~al.}(2012)\citenamefont{Dutra, Lourenco,
  S{\'a}Martins, Delfino, Stone, and Stevenson}}]{Dutra12}
\bibinfo{author}{\bibfnamefont{M.}~\bibnamefont{Dutra}},
  \bibinfo{author}{\bibfnamefont{O.}~\bibnamefont{Lourenco}},
  \bibinfo{author}{\bibfnamefont{J.~S.} \bibnamefont{S{\'a}Martins}},
  \bibinfo{author}{\bibfnamefont{A.}~\bibnamefont{Delfino}},
  \bibinfo{author}{\bibfnamefont{J.~R.} \bibnamefont{Stone}}, \bibnamefont{and}
  \bibinfo{author}{\bibfnamefont{P.~D.} \bibnamefont{Stevenson}},
  \bibinfo{journal}{Phys. Rev. C} \textbf{\bibinfo{volume}{85}},
  \bibinfo{pages}{035201} (\bibinfo{year}{2012}).

\bibitem[{\citenamefont{Dutra et~al.}(2014)\citenamefont{Dutra,
  Louren\ifmmode~\mbox{\c{c}}\else \c{c}\fi{}o, Avancini, Carlson, Delfino,
  Menezes, Provid\^encia, Typel, and Stone}}]{Dutra14}
\bibinfo{author}{\bibfnamefont{M.}~\bibnamefont{Dutra}},
  \bibinfo{author}{\bibfnamefont{O.}~\bibnamefont{Louren\ifmmode~\mbox{\c{c}}\else
  \c{c}\fi{}o}}, \bibinfo{author}{\bibfnamefont{S.~S.} \bibnamefont{Avancini}},
  \bibinfo{author}{\bibfnamefont{B.~V.} \bibnamefont{Carlson}},
  \bibinfo{author}{\bibfnamefont{A.}~\bibnamefont{Delfino}},
  \bibinfo{author}{\bibfnamefont{D.~P.} \bibnamefont{Menezes}},
  \bibinfo{author}{\bibfnamefont{C.}~\bibnamefont{Provid\^encia}},
  \bibinfo{author}{\bibfnamefont{S.}~\bibnamefont{Typel}}, \bibnamefont{and}
  \bibinfo{author}{\bibfnamefont{J.~R.} \bibnamefont{Stone}},
  \bibinfo{journal}{Phys. Rev. C} \textbf{\bibinfo{volume}{90}},
  \bibinfo{pages}{055203} (\bibinfo{year}{2014}).

\bibitem[{\citenamefont{Nik\ifmmode \check{s}\else
  \v{s}\fi{}i\ifmmode~\acute{c}\else \'{c}\fi{}
  et~al.}(2008)\citenamefont{Nik\ifmmode \check{s}\else
  \v{s}\fi{}i\ifmmode~\acute{c}\else \'{c}\fi{}, Vretenar, and
  Ring}}]{Niksic08}
\bibinfo{author}{\bibfnamefont{T.}~\bibnamefont{Nik\ifmmode \check{s}\else
  \v{s}\fi{}i\ifmmode~\acute{c}\else \'{c}\fi{}}},
  \bibinfo{author}{\bibfnamefont{D.}~\bibnamefont{Vretenar}}, \bibnamefont{and}
  \bibinfo{author}{\bibfnamefont{P.}~\bibnamefont{Ring}},
  \bibinfo{journal}{Phys. Rev. C} \textbf{\bibinfo{volume}{78}},
  \bibinfo{pages}{034318} (\bibinfo{year}{2008}).

\bibitem[{\citenamefont{Zhao et~al.}(2010)\citenamefont{Zhao, Li, Yao, and
  Meng}}]{Zhao10}
\bibinfo{author}{\bibfnamefont{P.~W.} \bibnamefont{Zhao}},
  \bibinfo{author}{\bibfnamefont{Z.~P.} \bibnamefont{Li}},
  \bibinfo{author}{\bibfnamefont{J.~M.} \bibnamefont{Yao}}, \bibnamefont{and}
  \bibinfo{author}{\bibfnamefont{J.}~\bibnamefont{Meng}},
  \bibinfo{journal}{Phys. Rev. C} \textbf{\bibinfo{volume}{82}},
  \bibinfo{pages}{054319} (\bibinfo{year}{2010}).

\bibitem[{\citenamefont{Dong et~al.}(2012)\citenamefont{Dong, Zuo, Gu, and
  Lombardo}}]{Dong12}
\bibinfo{author}{\bibfnamefont{J.}~\bibnamefont{Dong}},
  \bibinfo{author}{\bibfnamefont{W.}~\bibnamefont{Zuo}},
  \bibinfo{author}{\bibfnamefont{J.}~\bibnamefont{Gu}}, \bibnamefont{and}
  \bibinfo{author}{\bibfnamefont{U.}~\bibnamefont{Lombardo}},
  \bibinfo{journal}{Phys. Rev. C} \textbf{\bibinfo{volume}{85}},
  \bibinfo{pages}{034308} (\bibinfo{year}{2012}).

\bibitem[{\citenamefont{Brown et~al.}(2000)\citenamefont{Brown, Richter, and
  Lindsay}}]{Brown00}
\bibinfo{author}{\bibfnamefont{B.~A.} \bibnamefont{Brown}},
  \bibinfo{author}{\bibfnamefont{W.~A.} \bibnamefont{Richter}},
  \bibnamefont{and} \bibinfo{author}{\bibfnamefont{R.}~\bibnamefont{Lindsay}},
  \bibinfo{journal}{Phys. Lett. B} \textbf{\bibinfo{volume}{483}},
  \bibinfo{pages}{49} (\bibinfo{year}{2000}).

\bibitem[{\citenamefont{Mondal et~al.}(2015)\citenamefont{Mondal, Agrawal, and
  De}}]{Mondal15}
\bibinfo{author}{\bibfnamefont{C.}~\bibnamefont{Mondal}},
  \bibinfo{author}{\bibfnamefont{B.~K.} \bibnamefont{Agrawal}},
  \bibnamefont{and} \bibinfo{author}{\bibfnamefont{J.~N.} \bibnamefont{De}},
  \bibinfo{journal}{Phys. Rev. C} \textbf{\bibinfo{volume}{92}},
  \bibinfo{pages}{024302} (\bibinfo{year}{2015}).

\bibitem[{\citenamefont{Chen and Piekarewicz}(2015)}]{Chen15}
\bibinfo{author}{\bibfnamefont{W.-C.} \bibnamefont{Chen}} \bibnamefont{and}
  \bibinfo{author}{\bibfnamefont{J.}~\bibnamefont{Piekarewicz}},
  \bibinfo{journal}{Physics Letters B} \textbf{\bibinfo{volume}{748}},
  \bibinfo{pages}{284 } (\bibinfo{year}{2015}).

\bibitem[{\citenamefont{Todd-Rutel and Piekarewicz}(2005)}]{Todd-Rutel05}
\bibinfo{author}{\bibfnamefont{B.~G.} \bibnamefont{Todd-Rutel}}
  \bibnamefont{and}
  \bibinfo{author}{\bibfnamefont{J.}~\bibnamefont{Piekarewicz}},
  \bibinfo{journal}{Phys. Rev. Lett} \textbf{\bibinfo{volume}{95}},
  \bibinfo{pages}{122501} (\bibinfo{year}{2005}).

\bibitem[{\citenamefont{Furnstahl et~al.}(1997)\citenamefont{Furnstahl, Serot,
  and Tang}}]{Furnstahl97}
\bibinfo{author}{\bibfnamefont{R.}~\bibnamefont{Furnstahl}},
  \bibinfo{author}{\bibfnamefont{B.~D.} \bibnamefont{Serot}}, \bibnamefont{and}
  \bibinfo{author}{\bibfnamefont{H.-B.} \bibnamefont{Tang}},
  \bibinfo{journal}{Nucl. Phys.} \textbf{\bibinfo{volume}{A615}},
  \bibinfo{pages}{441} (\bibinfo{year}{1997}).

\bibitem[{\citenamefont{Boguta and Bodmer}(1977)}]{Boguta77}
\bibinfo{author}{\bibfnamefont{J.}~\bibnamefont{Boguta}} \bibnamefont{and}
  \bibinfo{author}{\bibfnamefont{A.~R.} \bibnamefont{Bodmer}},
  \bibinfo{journal}{Nucl. Phys.} \textbf{\bibinfo{volume}{A292}},
  \bibinfo{pages}{413} (\bibinfo{year}{1977}).

\bibitem[{\citenamefont{Boguta and Stoecker}(1983)}]{Boguta83}
\bibinfo{author}{\bibfnamefont{J.}~\bibnamefont{Boguta}} \bibnamefont{and}
  \bibinfo{author}{\bibfnamefont{H.}~\bibnamefont{Stoecker}},
  \bibinfo{journal}{Phys.Lett.} \textbf{\bibinfo{volume}{B120}},
  \bibinfo{pages}{289} (\bibinfo{year}{1983}).

\bibitem[{\citenamefont{Furnstahl}(2002)}]{Furnstahl02}
\bibinfo{author}{\bibfnamefont{R.}~\bibnamefont{Furnstahl}},
  \bibinfo{journal}{Nucl. Phys.} \textbf{\bibinfo{volume}{A706}},
  \bibinfo{pages}{85} (\bibinfo{year}{2002}).

\bibitem[{\citenamefont{Sil et~al.}(2005)\citenamefont{Sil, Centelles,
  Vi{\~n}as., and Piekarewicz}}]{Sil05}
\bibinfo{author}{\bibfnamefont{T.}~\bibnamefont{Sil}},
  \bibinfo{author}{\bibfnamefont{M.}~\bibnamefont{Centelles}},
  \bibinfo{author}{\bibfnamefont{X.}~\bibnamefont{Vi{\~n}as.}},
  \bibnamefont{and}
  \bibinfo{author}{\bibfnamefont{J.}~\bibnamefont{Piekarewicz}},
  \bibinfo{journal}{Phys. Rev.} \textbf{\bibinfo{volume}{C71}},
  \bibinfo{pages}{045502} (\bibinfo{year}{2005}).

\bibitem[{\citenamefont{Bevington}(1969)}]{Bevington69}
\bibinfo{author}{\bibfnamefont{P.~R.} \bibnamefont{Bevington}},
  \emph{\bibinfo{title}{Data Reduction and Error Analysis for\\ the Physical
  Sciences}} (\bibinfo{publisher}{McGraw-Hill, New York},
  \bibinfo{year}{1969}).

\bibitem[{\citenamefont{Wang et~al.}(2012)\citenamefont{Wang, Audi, Wapstra,
  Kondev, MacCormick, Xu, and Pfeiffer}}]{Wang12}
\bibinfo{author}{\bibfnamefont{M.}~\bibnamefont{Wang}},
  \bibinfo{author}{\bibfnamefont{G.}~\bibnamefont{Audi}},
  \bibinfo{author}{\bibfnamefont{A.}~\bibnamefont{Wapstra}},
  \bibinfo{author}{\bibfnamefont{F.}~\bibnamefont{Kondev}},
  \bibinfo{author}{\bibfnamefont{M.}~\bibnamefont{MacCormick}},
  \bibinfo{author}{\bibfnamefont{X.}~\bibnamefont{Xu}}, \bibnamefont{and}
  \bibinfo{author}{\bibfnamefont{B.}~\bibnamefont{Pfeiffer}},
  \bibinfo{journal}{Chinese Physics C} \textbf{\bibinfo{volume}{36}},
  \bibinfo{pages}{1603} (\bibinfo{year}{2012}).

\bibitem[{\citenamefont{Gallant et~al.}(2012)\citenamefont{Gallant, Bale,
  Brunner, Chowdhury, Ettenauer, Lennarz, Robertson, Simon, Chaudhuri, Holt
  et~al.}}]{Gallant12}
\bibinfo{author}{\bibfnamefont{A.~T.} \bibnamefont{Gallant}},
  \bibinfo{author}{\bibfnamefont{J.~C.} \bibnamefont{Bale}},
  \bibinfo{author}{\bibfnamefont{T.}~\bibnamefont{Brunner}},
  \bibinfo{author}{\bibfnamefont{U.}~\bibnamefont{Chowdhury}},
  \bibinfo{author}{\bibfnamefont{S.}~\bibnamefont{Ettenauer}},
  \bibinfo{author}{\bibfnamefont{A.}~\bibnamefont{Lennarz}},
  \bibinfo{author}{\bibfnamefont{D.}~\bibnamefont{Robertson}},
  \bibinfo{author}{\bibfnamefont{V.~V.} \bibnamefont{Simon}},
  \bibinfo{author}{\bibfnamefont{A.}~\bibnamefont{Chaudhuri}},
  \bibinfo{author}{\bibfnamefont{J.~D.} \bibnamefont{Holt}},
  \bibnamefont{et~al.}, \bibinfo{journal}{Phys. Rev. Lett.}
  \textbf{\bibinfo{volume}{109}}, \bibinfo{pages}{032506}
  (\bibinfo{year}{2012}).

\bibitem[{\citenamefont{Wienholtz and {\it et. al.}}(2013)}]{Wienholtz13}
\bibinfo{author}{\bibfnamefont{F.}~\bibnamefont{Wienholtz}} \bibnamefont{and}
  \bibinfo{author}{\bibnamefont{{\it et. al.}}}, \bibinfo{journal}{Nature}
  \textbf{\bibinfo{volume}{498}}, \bibinfo{pages}{346} (\bibinfo{year}{2013}).

\bibitem[{\citenamefont{Angeli and Marinova}(2013)}]{Angeli13}
\bibinfo{author}{\bibfnamefont{I.}~\bibnamefont{Angeli}} \bibnamefont{and}
  \bibinfo{author}{\bibfnamefont{K.}~\bibnamefont{Marinova}},
  \bibinfo{journal}{Atomic Data and Nuclear Data Tables}
  \textbf{\bibinfo{volume}{99}}, \bibinfo{pages}{69 } (\bibinfo{year}{2013}).

\bibitem[{\citenamefont{Demorest et~al.}(2010)\citenamefont{Demorest, Pennucci,
  Ransom, Roberts, and Hessels}}]{Demorest10}
\bibinfo{author}{\bibfnamefont{P.~B.} \bibnamefont{Demorest}},
  \bibinfo{author}{\bibfnamefont{T.}~\bibnamefont{Pennucci}},
  \bibinfo{author}{\bibfnamefont{S.~M.} \bibnamefont{Ransom}},
  \bibinfo{author}{\bibfnamefont{M.~S.~E.} \bibnamefont{Roberts}},
  \bibnamefont{and} \bibinfo{author}{\bibfnamefont{J.~W.~T.}
  \bibnamefont{Hessels}}, \bibinfo{journal}{Nature}
  \textbf{\bibinfo{volume}{467}}, \bibinfo{pages}{1081} (\bibinfo{year}{2010}).

\bibitem[{\citenamefont{Antoniadis and {\it et. al}}(2013)}]{Antoniadis13}
\bibinfo{author}{\bibfnamefont{J.}~\bibnamefont{Antoniadis}} \bibnamefont{and}
  \bibinfo{author}{\bibnamefont{{\it et. al}}}, \bibinfo{journal}{Science}
  \textbf{\bibinfo{volume}{340}}, \bibinfo{pages}{448} (\bibinfo{year}{2013}).

\bibitem[{\citenamefont{Kl\"upfel et~al.}(2009)\citenamefont{Kl\"upfel,
  Reinhard, B\"urvenich, and Maruhn}}]{Klupfel09}
\bibinfo{author}{\bibfnamefont{P.}~\bibnamefont{Kl\"upfel}},
  \bibinfo{author}{\bibfnamefont{P.-G.} \bibnamefont{Reinhard}},
  \bibinfo{author}{\bibfnamefont{T.~J.} \bibnamefont{B\"urvenich}},
  \bibnamefont{and} \bibinfo{author}{\bibfnamefont{J.~A.}
  \bibnamefont{Maruhn}}, \bibinfo{journal}{Phys. Rev. C}
  \textbf{\bibinfo{volume}{79}}, \bibinfo{pages}{034310}
  (\bibinfo{year}{2009}).

\bibitem[{\citenamefont{Horowitz and Piekarewicz}(2001)}]{Horowitz01a}
\bibinfo{author}{\bibfnamefont{C.~J.} \bibnamefont{Horowitz}} \bibnamefont{and}
  \bibinfo{author}{\bibfnamefont{J.}~\bibnamefont{Piekarewicz}},
  \bibinfo{journal}{Phys. Rev. C} \textbf{\bibinfo{volume}{64}},
  \bibinfo{pages}{062802(R)} (\bibinfo{year}{2001}).

\bibitem[{\citenamefont{Wang and Chen}(2015)}]{Wang15}
\bibinfo{author}{\bibfnamefont{R.}~\bibnamefont{Wang}} \bibnamefont{and}
  \bibinfo{author}{\bibfnamefont{L.-W.} \bibnamefont{Chen}},
  \bibinfo{journal}{Phys. Rev. C} \textbf{\bibinfo{volume}{92}},
  \bibinfo{pages}{031303} (\bibinfo{year}{2015}).

\bibitem[{\citenamefont{Chen et~al.}(2005)\citenamefont{Chen, Ko, and
  Li}}]{Chen05}
\bibinfo{author}{\bibfnamefont{L.-W.} \bibnamefont{Chen}},
  \bibinfo{author}{\bibfnamefont{C.~M.} \bibnamefont{Ko}}, \bibnamefont{and}
  \bibinfo{author}{\bibfnamefont{B.-A.} \bibnamefont{Li}},
  \bibinfo{journal}{Phys. Rev. C} \textbf{\bibinfo{volume}{72}},
  \bibinfo{pages}{064309} (\bibinfo{year}{2005}).

\bibitem[{\citenamefont{Kortelainen et~al.}(2010)\citenamefont{Kortelainen,
  Lesinski, Mor\'e, Nazarewicz, Sarich, Schunck, Stoitsov, and
  Wild}}]{Kortelainen10}
\bibinfo{author}{\bibfnamefont{M.}~\bibnamefont{Kortelainen}},
  \bibinfo{author}{\bibfnamefont{T.}~\bibnamefont{Lesinski}},
  \bibinfo{author}{\bibfnamefont{J.}~\bibnamefont{Mor\'e}},
  \bibinfo{author}{\bibfnamefont{W.}~\bibnamefont{Nazarewicz}},
  \bibinfo{author}{\bibfnamefont{J.}~\bibnamefont{Sarich}},
  \bibinfo{author}{\bibfnamefont{N.}~\bibnamefont{Schunck}},
  \bibinfo{author}{\bibfnamefont{M.~V.} \bibnamefont{Stoitsov}},
  \bibnamefont{and} \bibinfo{author}{\bibfnamefont{S.}~\bibnamefont{Wild}},
  \bibinfo{journal}{Phys. Rev. C} \textbf{\bibinfo{volume}{82}},
  \bibinfo{pages}{024313} (\bibinfo{year}{2010}).

\bibitem[{\citenamefont{Dobaczewski et~al.}(2014)\citenamefont{Dobaczewski,
  Nazarewicz, and Reinhard}}]{Dobaczewski14}
\bibinfo{author}{\bibfnamefont{J.}~\bibnamefont{Dobaczewski}},
  \bibinfo{author}{\bibfnamefont{W.}~\bibnamefont{Nazarewicz}},
  \bibnamefont{and} \bibinfo{author}{\bibfnamefont{P.-G.}
  \bibnamefont{Reinhard}}, \bibinfo{journal}{J Phys. G}
  \textbf{\bibinfo{volume}{41}}, \bibinfo{pages}{074001}
  (\bibinfo{year}{2014}).

\bibitem[{\citenamefont{Vretenar et~al.}(2003)\citenamefont{Vretenar,
  Nik$\check{s}$i$\grave{c}$, and Ring}}]{Vretenar03}
\bibinfo{author}{\bibfnamefont{D.}~\bibnamefont{Vretenar}},
  \bibinfo{author}{\bibfnamefont{T.}~\bibnamefont{Nik$\check{s}$i$\grave{c}$}},
  \bibnamefont{and} \bibinfo{author}{\bibfnamefont{P.}~\bibnamefont{Ring}},
  \bibinfo{journal}{Phys. Rev. C} \textbf{\bibinfo{volume}{68}},
  \bibinfo{pages}{024310} (\bibinfo{year}{2003}).

\bibitem[{\citenamefont{Nik$\check{s}$i$\grave{c}$
  et~al.}(2014)\citenamefont{Nik$\check{s}$i$\grave{c}$, Paar, Vretenar, and
  Ring}}]{Niksic14}
\bibinfo{author}{\bibfnamefont{T.}~\bibnamefont{Nik$\check{s}$i$\grave{c}$}},
  \bibinfo{author}{\bibfnamefont{N.}~\bibnamefont{Paar}},
  \bibinfo{author}{\bibfnamefont{D.}~\bibnamefont{Vretenar}}, \bibnamefont{and}
  \bibinfo{author}{\bibfnamefont{P.}~\bibnamefont{Ring}},
  \bibinfo{journal}{Computer Physics Communications}
  \textbf{\bibinfo{volume}{185}}, \bibinfo{pages}{1808 }
  (\bibinfo{year}{2014}).

\bibitem[{\citenamefont{Lalazissis et~al.}(1999)\citenamefont{Lalazissis,
  Vretenar, Ring, Stoitsov, and Robledo}}]{Lalazissis99a}
\bibinfo{author}{\bibfnamefont{G.~A.} \bibnamefont{Lalazissis}},
  \bibinfo{author}{\bibfnamefont{D.}~\bibnamefont{Vretenar}},
  \bibinfo{author}{\bibfnamefont{P.}~\bibnamefont{Ring}},
  \bibinfo{author}{\bibfnamefont{M.}~\bibnamefont{Stoitsov}}, \bibnamefont{and}
  \bibinfo{author}{\bibfnamefont{L.~M.} \bibnamefont{Robledo}},
  \bibinfo{journal}{Phys. Rev. C} \textbf{\bibinfo{volume}{60}},
  \bibinfo{pages}{014310} (\bibinfo{year}{1999}).

\bibitem[{\citenamefont{De et~al.}(2015)\citenamefont{De, Samaddar, and
  Agrawal}}]{De15}
\bibinfo{author}{\bibfnamefont{J.~N.} \bibnamefont{De}},
  \bibinfo{author}{\bibfnamefont{S.~K.} \bibnamefont{Samaddar}},
  \bibnamefont{and} \bibinfo{author}{\bibfnamefont{B.~K.}
  \bibnamefont{Agrawal}}, \bibinfo{journal}{Phys. Rev. C}
  \textbf{\bibinfo{volume}{92}}, \bibinfo{pages}{014304}
  (\bibinfo{year}{2015}).

\bibitem[{\citenamefont{Zhang and Chen}(2013)}]{Zhang13}
\bibinfo{author}{\bibfnamefont{Z.}~\bibnamefont{Zhang}} \bibnamefont{and}
  \bibinfo{author}{\bibfnamefont{L.-W.} \bibnamefont{Chen}},
  \bibinfo{journal}{Physics Letters B} \textbf{\bibinfo{volume}{726}},
  \bibinfo{pages}{234 } (\bibinfo{year}{2013}).

\bibitem[{\citenamefont{Zhang and Chen}(2015)}]{Zhang15}
\bibinfo{author}{\bibfnamefont{Z.}~\bibnamefont{Zhang}} \bibnamefont{and}
  \bibinfo{author}{\bibfnamefont{L.-W.} \bibnamefont{Chen}},
  \bibinfo{journal}{Phys. Rev. C} \textbf{\bibinfo{volume}{92}},
  \bibinfo{pages}{031301} (\bibinfo{year}{2015}).

\end{thebibliography}

\end{document}